\documentclass[a4paper]{article}

\usepackage{amsfonts}
\usepackage{graphicx}
\usepackage{url}
\usepackage{amsthm}
\usepackage{color}
\usepackage{braket}
\usepackage{extarrows}
\usepackage{titling}
\usepackage{amsmath}
\usepackage{hyperref}

\date{}

\title{Secure $N$-dimensional simultaneous dense coding and applications}

\author{Haozhen Situ\textsuperscript{1,2}\thanks{situhaozhen@gmail.com} \and Daowen Qiu\textsuperscript{1,3}\thanks{issqdw@mail.sysu.edu.cn} \and Paulo Mateus\textsuperscript{4,5}\thanks{pmat@math.ist.utl.pt} \and Nikola Paunkovi\'c\textsuperscript{4,5}\thanks{npaunkovic@gmail.com}}

\begin{document}
\maketitle

\noindent\textsuperscript{1} Department of Computer Science, Sun Yat-sen University, Guangzhou 510006, China\\
\textsuperscript{2} College of Mathematics and Informatics, South China Agricultural University, Guangzhou 510642, China\\
\textsuperscript{3} The Guangdong Key Laboratory of Information Security Technology, Sun Yat-sen University, Guangzhou 510006, China\\
\textsuperscript{4} Departamento de Matem\' atica, Instituto Superior T\' ecnico, Universidade de Lisboa, Av. Rovisco Pais 1049-001, Lisboa, Portugal\\
\textsuperscript{5} SQIG -- Security and Quantum Information Group, Instituto de Telecomunica\c{c}\~{o}es, Av. Rovisco Pais 1049-001, Lisboa, Portugal

%

\begin{abstract}
Simultaneous dense coding guarantees that Bob and Charlie
simultaneously receive their respective information from Alice in
their respective processes of dense coding. The idea is to use the so-called locking operation to
``lock'' the entanglement channels, thus requiring a joint unlocking
operation by Bob and Charlie in order to simultaneously obtain the
information sent by Alice. We present some new results on simultaneous dense coding: (1) We
propose three simultaneous dense coding protocols, which use
different $N$-dimensional entanglement (Bell state, W
state and GHZ state). (2) Besides the quantum Fourier transform, two
new locking operators are introduced (the double controlled-NOT operator and
the SWAP operator). (3) In the case that spatially distant Bob and
Charlie have to finalise the protocol by implementing the unlocking
operation through communication, we improve our protocol's fairness, with
respect to  Bob and Charlie, by
implementing the unlocking operation in series of steps. (4) We
improve the security of simultaneous dense coding against the
intercept-resend attack. (5) We show that simultaneous dense coding
can be used to implement a fair contract signing protocol. (6) We
also show that the $N$-dimensional quantum Fourier transform can act
as the locking operator in simultaneous teleportation of
$N$-level quantum systems.
\end{abstract}


\section{Introduction}
Due to the Holevo bound~\cite{H73}, at most $\log_2N$ bits of information can be transmitted via a
qu$N$it ($N$-level quantum system). Dense coding, proposed by Bennett and Wiesner~\cite{BW92} in 1992, increases the classical capacity of a quantum
communication channel with the help of prior entanglement~\cite{HHHH09}. If the sender and the receiver share a pair of
entangled qu$N$its, $2\log_2N$ bits can be transmitted via a qu$N$it~\cite{BW92}.

In the simplest case of dense coding two parties, Alice and Bob,
share a pair of entangled qubits (2-level quantum systems)
in a Bell state. Alice first performs one of the four local
unitary operations $I$, $\sigma_x$, $i\sigma_y$ and $\sigma_z$
(where $\sigma_j$ are the Pauli matrices) on her qubit to encode 2
bits of information, transforming the entangled pair into one of the
four mutually orthogonal Bell states. Then Alice sends her qubit to
Bob through a quantum channel. Bob is now able to measure both
qubits in the Bell basis to obtain one of the four possible outcomes
correlated with the operations performed by Alice. Thus, Alice can
transmit 2 bits of information to Bob by manipulating and sending
only one qubit.

Thus far, dense coding has been extensively studied in various ways.
For example, dense coding that utilises high-dimensional entangled
states has been studied in~\cite{LLTL02,WG02,FZ06}, non-maximally
entanglement channels in~\cite{BE95,HJSWW96,ZB00,HLG00,B01,PPA05,MOR05,WCSG06,BGMW08,BC09,GWB09,HYLH09,S13a},
while multipartite entanglement channels have also
been considered in~\cite{BVK98,LAH02,YC06,BLSSDM06,AP06,WSXL07,LQ07,MP08}. Another generalisation is to perform the
communication task under the control of a third party, so-called
controlled dense coding~\cite{HLG01,LO09,HL09,S13b}.

Inspired by the simultaneous teleportation scheme proposed by Wang
{\it et al}~\cite{WYGL08}, we have proposed a simultaneous dense
coding (SDC) scheme~\cite{SQ10}, which guarantees that Bob
and Charlie (the receivers) simultaneously receive their respective
information from Alice (the sender) in their respective processes of
dense coding. In this scheme, Alice first performs a locking
operation to entangle the particles from two independent quantum
entanglement channels, and therefore the receivers cannot obtain
their respective information separately, before performing the unlocking
operation together. The quantum Fourier transform and its inverse
are used as the locking and unlocking operators, respectively.

Simultaneous dense coding may be relevant and useful in various applications in which Bob and Charlie can be either close or separated. If Alice has two different secrets, one for Bob and another for Charlie, she can utilise simultaneous dense coding to guarantee that Bob and Charlie simultaneously reveal their respective secrets. Bob does not know Charlie's secret and vice versa. For example, boss wants two employees to simultaneously carry out two confidential commercial activities under the condition that the sensitive information of each activity is only revealed to
whom is in charge of that activity.

There are other applications improving some models or tasks of quantum communication. In Sec.~\ref{application}, we show that simultaneous dense coding can be used to implement a fair contract signing protocol~\cite{EY80}
between spatially distant Bob and Charlie. In this case, Bob and Charlie are separated so that there is a problem of how to implement the unlocking operation fairly. We discuss this problem in Sec.\ref{fairness}.

In this paper, we give some new results on simultaneous dense coding
and teleportation. The main improvements over earlier proposals are: (i) introduction of new locking operations (the double-controlled-NOT and the SWAP operators) for SDC schemes that were not used before, showing that the protocol can be achieved by using three different $N$-dimensional quantum states (Bell, W and GHZ states) that were not mentioned in that context in the literature before; (ii) designing a strategy which achieves the fairness of SDC, allowing for the protocol to be performed even when mutually mistrustful Bob and Charlie are situated on spatially distant locations, a feature which previous SDC proposals do not allow; (iii) designing a strategy which also achieves the security of SDC against intercept-resend attack, improving the security of SDC over the previous proposals; (iv) introducing simultaneous teleportation scheme for the transfer of arbitrary number of $N$-level quantum states. In addition to that, we proposed contract signing protocol based on a fair and secure SDC protocol. Having more options for achieving simultaneous transmission of (densely coded) information presents opportunities for designing SDC protocols with improved security features, and in addition provides wider range of possibly suitable experimental realisations. Indeed, achieving the fairness of SDC scheme, a feature not satisfied by previous SDC proposals, is only possible with the double-controlled-NOT and the SWAP locking operator. Also, the security of the protocol against intercept-resend attack is a novel feature of SDC schemes introduced in our paper.

The paper is organised as follows. In Sec. \ref{n-dimension}, we introduce the
locking operators (i.e. the quantum Fourier transform, the double
controlled-NOT operator and the SWAP operator) and propose three
simultaneous dense coding protocols utilising different
$N$-dimensional quantum states (i.e., Bell states, W states, and GHZ
states). In Sec. \ref{fairness}, we improve the fairness of the protocol by
implementing the unlocking operation between spatially distant Bob
and Charlie. In Sec. \ref{security}, we improve the security of
simultaneous dense coding against the intercept-resend attack.
In Sec.~\ref{application}, we show that simultaneous dense coding can be
used to implement a fair contract signing protocol. In Sec.
\ref{teleportation}, we show that the $N$-dimensional quantum
Fourier transform can act as the locking operator in simultaneous
teleportation of qu$N$its. A brief conclusion follows in Sec.
\ref{conclusion}.

\section{\label{n-dimension} $N$-dimensional simultaneous dense coding}

A qu$N$it is an $N$-dimensional quantum system. States of
$N$-dimensional quantum systems can be mapped onto $n = \log_2 (N)$
qubits, and throughout the paper we will assume that a qu$N$it is
realised as an ordered array of qubits. We also assume that the
dimension $N$ of qu$N$its satisfies the requirement $N=2^n$, with
$n\in \mathbb{N}$. In other words, one qu$N$it consists of (or can
be mapped to) $n$ qubits.

In the task of $N$-dimensional simultaneous dense coding (SDC),
Alice intends to use dense coding to send two c$N$its (arrays of
classical bits) $b_1,b_2\in \{0,1,\ldots , N-1\}$ to Bob and two
c$N$its $c_1,c_2\in \{0,1,\ldots , N-1\}$ to Charlie, under the
condition that Bob and Charlie must collaborate to simultaneously
find out what she sends.

In the following subsections, we propose three protocols, one using $N$-dimensional Bell states, the other using W states, and the third using GHZ states, as the entanglement channels, respectively. The idea behind these protocols is
to perform the locking operator on Alice's qu$N$its before sending
them to Bob and Charlie. After receiving Alice's qu$N$its, the
states of Bob's subsystem and Charlie's subsystem are independent of
$(b_1,b_2)$ and $(c_1,c_2)$, respectively, so that they know nothing about the
encoded information. Only after performing the unlocking operator
(the inverse of the locking operator) together, Bob and Charlie can obtain
$(b_1,b_2)$ and $(c_1,c_2)$, respectively.

In this section, we assume that Bob and Charlie are at the same
site. In order to implement the unlocking operator, they can input
their particles into a physical device that can unlock the
particles, and then get their respective output particles. The problem of how distant Bob and Charlie can perform the unlocking
operator fairly is discussed in Sec.~\ref{fairness}.

\subsection{DCNOT locking operator}
Before introducing the protocols, let us first have a look at the
DCNOT (double controlled-NOT) operator, which is used as the locking
operator. The DCNOT operator is composed of two CNOT
(controlled-NOT) operators. The CNOT operator has two input qubits.
If the first qubit (the control qubit) is in state $\ket 1$, CNOT flips the state of the second
qubit (the target qubit). If the control qubit is in state $\ket 0$, CNOT does
nothing to the target qubit ($\ket 0$ and $\ket 1$ are the basis vectors). Its action can be described as ($x,y = 0,1$ are bit values):
\begin{align}
|x\rangle |y\rangle \rightarrow |x\rangle |x\oplus y\rangle,
\end{align}
where $\oplus$ denotes bitwise addition modulo $2$.

The DCNOT operator is formed by performing the first CNOT with the first qubit as the control and the second being the target, and then a second CNOT with inverse roles of the qubits (the first qubit as the target and the second being the control). Its action can be
described as:
\begin{align}
|x\rangle |y\rangle \rightarrow |y\rangle |x\oplus y\rangle,
\end{align}
and the inverse DCNOT operator can be described as:
\begin{align}
|x\rangle |y\rangle \rightarrow |x\oplus y\rangle |x\rangle.
\end{align}

Our qu$N$its $|x\rangle_{A_1} |y\rangle_{A_2}$, with $x,y \in \{0,1,\ldots, N-1\}$, are arrays of $n$
qubits, and can be written as:
\begin{equation}
|x_1,\ldots ,x_n\rangle_{A_{1,1}\ldots A_{1,n}} |y_1,\ldots,
y_n\rangle_{A_{2,1}\ldots A_{2,n}},
\end{equation}
where $x_1\ldots x_n$ and $y_1\ldots y_n$ are binary representations of $x$ and $y$, respectively, and $A_{i,j}$ denote different qubits ($i =1,2$ and $j = 1,2,\ldots N$). We define the $N$-dimensional DCNOT operator by applying the DCNOT
operator on each pair of $A_{1,j}A_{2,j}$:

\begin{align}
|x_1,\ldots ,x_n\rangle_{A_{1,1}\ldots A_{1,n}} & |y_1,\ldots,
y_n\rangle_{A_{2,1}\ldots A_{2,n}}  
\rightarrow  |y_1,\ldots ,y_n\rangle_{A_{1,1}\ldots A_{1,n}}
|x_1\oplus y_1,\ldots, x_n \oplus y_n\rangle_{A_{2,1}\ldots
A_{2,n}}, 
\end{align}

that is ($A_1 = A_{1,1}A_{1,2}\ldots A_{1,N}$ and analogously for $A_2$),

\begin{align}
|x\rangle_{A_1} |y\rangle_{A_2} \rightarrow |y\rangle_{A_1} |x\oplus
y\rangle_{A_2}.
\end{align}

Analogously, by applying the inverse DCNOT operator on each pair of
$A_{1,j}A_{2,j}$, we have the $N$-dimensional inverse DCNOT
operator:
\begin{align}
|x_1,\ldots ,x_n\rangle_{A_{1,1}\ldots A_{1,n}} & |y_1,\ldots,
y_n\rangle_{A_{2,1}\ldots A_{2,n}}  
 \rightarrow |x_1\oplus y_1,\ldots ,x_n\oplus
y_n\rangle_{A_{1,1}\ldots A_{1,n}} |x_1,\ldots,
x_n\rangle_{A_{2,1}\ldots A_{2,n}}, 
\end{align}
that is,
\begin{align}
|x\rangle_{A_1} |y\rangle_{A_2} \rightarrow |x\oplus y\rangle_{A_1}
|x\rangle_{A_2}.
\end{align}

\subsection{SDC Protocol 1: using $N$-dimensional Bell state}
Protocol 1 uses the following $N$-dimensional Bell basis~\cite{BBCJPW93}, for composite systems of two qu$N$its $1$ and $2$:
\begin{align}
|\phi (xy)\rangle_{12} = \frac{1}{\sqrt{N}} \sum_{j=0}^{N-1}
e^{\frac{2\pi i}{N}jx} |j + y\rangle_1 |j\rangle_2,
\end{align}
where $x,y \in \{0,1,\ldots, N-1\}$. In the rest of the paper,
addition and subtraction inside kets are
done modulo $N$.

The local (acting onto system $1$ only) unitary operators $U(xy)$ transform $|\phi (00)\rangle_{12}$
into $|\phi (xy)\rangle_{12}$:
\begin{align}
\Big( U(xy)\otimes I\Big) |\phi(00)\rangle_{12} =
|\phi(xy)\rangle_{12},
\end{align}
where $U(xy)=X^yZ^x$, $X: |j\rangle \longrightarrow |j + 1\rangle$
is the shift operator, and $Z: |j\rangle \longrightarrow e^{\frac{2\pi
i}{N}j}|j\rangle$ is the rotation operator.

The set $\{ |\phi (xy)\rangle_{12} \}$ forms an orthonormal basis of
completely distinguishable states, and for that reason each of its
elements can be used to carry information $(xy)$. The unitary
operators $U(xy)$ are then used to encode that information $(xy)$
into the initial state $|\phi (00)\rangle_{12}$.

In the initialisation phase of Protocol 1, Alice, Bob and Charlie
share two pairs of entangled qu$N$its $|\phi(00)\rangle_{A_1B}$ and
$|\phi(00)\rangle_{A_2C}$, where subscripts $A_1$ and $A_2$ denote
Alice's two qu$N$its, subscript $B$ denotes Bob's qu$N$it and
subscript $C$ denotes Charlie's qu$N$it. The initial quantum state
of the composite system is (note that for reasons of simplicity, we
drop the subscript $A_1BA_2C$ for $\Omega$-states
$|\Omega(0)\rangle$ etc.)
\begin{align}
|\Omega(0)\rangle = |\phi(00)\rangle_{A_1B}\otimes
|\phi(00)\rangle_{A_2C}.
\end{align}

Protocol 1 consists of five steps:

(1) {\it Encoding.} Alice performs unitary operators $U(b_1b_2)$ on
qu$N$it $A_1$ and $U(c_1c_2)$ on qu$N$it $A_2$ to encode $(b_1,
b_2)$ and $(c_1, c_2)$, respectively, like in the original dense coding
scheme~\cite{BW92}. After that, the state of the composite system
becomes
\begin{align}|\Omega(1)\rangle = U_{A_1}(b_1b_2)\otimes
U_{A_2}(c_1c_2)|\Omega(0)\rangle =|\phi(b_1b_2)\rangle_{A_1B}
\otimes |\phi(c_1c_2)\rangle_{A_2C}.
\end{align}

(2) {\it Locking.} Alice performs the DCNOT operator on qu$N$its
$A_1A_2$ to lock the entanglement channels. The state of the
composite system becomes
\begin{align}
|\Omega(2)\rangle = DCNOT_{A_1A_2}\Big( |\phi(b_1b_2)\rangle_{A_1B}
\otimes |\phi(c_1c_2)\rangle_{A_2C}\Big).
\end{align}

(3) {\it Communication.} Alice sends qu$N$it $A_1$ to Bob and
qu$N$it $A_2$ to Charlie, like the original dense coding scheme~\cite{BW92}.

(4) {\it Unlocking.} Bob and Charlie collaborate to perform the
inverse $DCNOT$ operator on qu$N$its $A_1A_2$. The state of the
composite system becomes
\begin{align}
|\Omega(3)\rangle & = DCNOT_{A_1A_2}^\dagger|\Omega(2)\rangle
=|\phi(b_1b_2)\rangle_{A_1B} \otimes |\phi(c_1c_2)\rangle_{A_2C}.
\end{align}

(5) {\it Decoding.} Bob and Charlie measure qu$N$its $A_1B$ and
qu$N$its $A_2C$ in the $N$-dimensional Bell basis respectively to
obtain $(b_1,b_2)$ and $(c_1,c_2)$, like the original dense coding
scheme~\cite{BW92}.

The following theorem demonstrates the validity of Protocol 1.

\bigskip

{\bf Theorem 1.} {\it Neither Bob nor Charlie alone can learn the
encoded information from the states of their subsystems before step
4 (Unlocking) of Protocol 1.}

\bigskip

\emph{Proof} After step 2 (\emph{Locking}), the state of the
composite system becomes
\begin{align}
|\Omega(2)\rangle = & DCNOT_{A_1A_2}\Bigg(
\frac{1}{\sqrt{N}}\sum_{j=0}^{N-1}e^{\frac{2\pi i}{N} jb_1}|j +
b_2\rangle_{A_1}|j\rangle_{B}
 \otimes \frac{1}{\sqrt{N}}\sum_{k=0}^{N-1}e^{\frac{2\pi
i}{N}kc_1}|k + c_2\rangle_{A_2}|k\rangle_{C}\Bigg)
\nonumber\\
= & \frac{1}{N} \sum_{j,k=0}^{N-1} e^{\frac{2\pi i}{N}(jb_1+kc_1)}
|k+c_2\rangle_{A_1} |(j + b_2) \oplus (k +
c_2)\rangle_{A_2}|j\rangle_{B} |k\rangle_{C},
\end{align}

and the reduced density matrix in subsystem $A_1B$ is
\begin{align}
\rho_{A_1B} & = \sum_{j,k=0}^{N-1} (_{A_2}\langle j|_{C}\langle
k|)|\Omega(2)\rangle \langle\Omega(2)|
(|j\rangle_{A_2} |k\rangle_{C})\nonumber\\
&  = \frac{1}{N^2} \sum_{j,k=0}^{N-1} |j\rangle_{A_1}
|k\rangle_{BA_1}\langle j|_{B}\langle k|\nonumber\\
&  = I_{A_1B}/N^2.
\end{align}

We can calculate the reduced density matrix in subsystem $A_2C$ in
the same way and get $\rho_{A_2C}=I_{A_2C}/N^2$. Because
$\rho_{A_1B}$ and $\rho_{A_2C}$ are independent of $(b_1, b_2)$ and
$(c_1, c_2)$, Bob and Charlie know nothing about the encoded
information before step 4 (\emph{Unlocking}). $\Box$

\subsection{SDC Protocol 2: using $N$-dimensional W state}
Li and Qiu~\cite{LQ07} presented a sufficient and necessary
condition for a $N$-dimensional W state to be suitable for perfect
teleportation and dense coding, and then they generalised the states
of W-class to multi-particle systems with $N$-dimension:
\begin{align}
|W(00)\rangle_{123} = & \frac{1}{\sqrt{N}} \Big( \frac{1}{\sqrt{2}}
\sum_{j=1}^{N-1} |j-1\rangle_1 (|0j\rangle+|j0\rangle)_{23}
 + |N-1\rangle_1 |00\rangle_{23}\Big).
\end{align}

Alice uses unitary operators
\begin{align}
U(xy)=\sum_{j=0}^{N-1}e^{\frac{2\pi i}{N}(j - y)x}|j -
y\rangle\langle j|
\end{align} to encode her information:
\begin{align}|W(xy)\rangle_{123} = \Big( U(xy) \otimes I \otimes I\Big) |W(00)\rangle_{123},\end{align} where
$x,y \in \{0,1,\ldots, N-1\}$.

In the initialisation phase of Protocol 2, Alice, Bob and Charlie
share two pairs of entangled qu$N$its $|W(00)\rangle_{A_1B}$ and
$|W(00)\rangle_{A_2C}$, where subscript $A_1$ and $A_2$ denote
Alice's two qu$N$its, subscript $B$ denotes Bob's two qu$N$its and
subscript $C$ denotes Charlie's two qu$N$its. Protocol 2 consists of
five steps:

(1) {\it Encoding.} Alice performs unitary operators $U(b_1b_2)$ on
qu$N$it $A_1$ and $U(c_1c_2)$ on qu$N$it $A_2$ to encode $(b_1,
b_2)$ and $(c_1, c_2)$, respectively.

(2) {\it Locking.} Alice performs the DCNOT operator on qu$N$its
$A_1A_2$.

(3) {\it Communication.} Alice sends qu$N$it $A_1$ to Bob and
qu$N$it $A_2$ to Charlie.

(4) {\it Unlocking.} Bob and Charlie collaborate to perform the
inverse $DCNOT$ operator on qu$N$its $A_1A_2$.

(5) {\it Decoding.} Bob and Charlie make the von Neumann measurement
using the orthogonal states $\{|W(xy)\rangle\}$ on qu$N$its $A_1B$
and qu$N$its $A_2C$, respectively, to obtain $(b_1,b_2)$ and
$(c_1,c_2)$.

The following theorem demonstrates the validity of Protocol 2.

\newpage

{\bf Theorem 2.} {\it Neither Bob nor Charlie alone can learn the
encoded information from the states of their subsystems before step
4 (Unlocking) of Protocol 2.}

\bigskip

\emph{Proof} The proof is similar to that of Protocol 1. After step
2 (\emph{Locking}), the reduced density matrix
\begin{align}
\rho_{A_1B} & =  \frac{1}{N^2} \sum_{j=0}^{N-1}
|j\rangle_{A_1}|00\rangle_{BA_1}
\langle j| _{B}\langle 00| + \frac{1}{2N^2} \sum_{k=1}^{N-1}\sum_{j=0}^{N-1}
|j\rangle_{A_1}(|0k\rangle+|k0\rangle)_{BA_1} \langle j|
_{B}(\langle 0k|+\langle k0|) \nonumber \\
& =  \frac{1}{N} \left[ \sum_{j=0}^{N-1} |j\rangle_{A_1} \langle
j|\right]  \otimes \frac{1}{N} \left[ |00\rangle_{B}\langle 00| +
\frac{1}{2}\sum_{k=1}^{N-1} ( |0k\rangle+|k0\rangle )_{B}( \langle
0k|+\langle k0| ) \right],
\end{align}
and analogously for $rho_{A_2C}$. Because $\rho_{A_1B}$ and $\rho_{A_2C}$ are independent of $(b_1,
b_2)$ and $(c_1, c_2)$, Bob and Charlie know nothing about the
encoded information before step 4 (\emph{Unlocking}). $\Box$

\subsection{SDC Protocol 3: using $N$-dimensional GHZ state}
Protocol 3 uses the following $N$-dimensional GHZ states~\cite{LLTL02}:
\begin{align}
|GHZ(xy)\rangle_{123} = \frac{1}{\sqrt{N}} \sum_{j=0}^{N-1}
e^{\frac{2\pi i}{N}jx} |j + y\rangle_1 |jj\rangle_{23},
\end{align}
where $x,y \in \{0,1,\ldots, N-1\}$.

The unitary operators $U(xy)$ transform $|GHZ(00)\rangle_{123}$
into $|GHZ(xy)\rangle_{123}$:
\begin{align}
\Big( U(xy) \otimes I \otimes I \Big) |GHZ(00)\rangle_{123} =
|GHZ(xy)\rangle_{123},
\end{align}
where $U(xy)=X^yZ^x$, $X: |j\rangle \longrightarrow |j + 1\rangle$
is the shift operator, $Z: |j\rangle \longrightarrow e^{\frac{2\pi
i}{N}j}|j\rangle$ is the rotation operator.

Protocol 3 is similar to Protocols 1 and 2, and the proof of its validity is
analogous to those of Protocols 1 and 2.

\subsection{Other locking operators}
In this subsection, we introduce another two locking operators: the
quantum Fourier transform and the SWAP operator.

The two-qu$N$it quantum Fourier transform is defined by
\begin{align}
|x\rangle_{A_1}|y\rangle_{A_2}\rightarrow\frac{1}{N}\sum_{j=0}^{N-1}\sum_{k=0}^{N-1}e^{\frac{2\pi
i}{N^2}(xN+y)(jN+k)}|j\rangle_{A_1}|k\rangle_{A_2}.
\end{align}

The SWAP operator simply swaps two qubits (or qu$N$its):
\begin{align}
|\varphi\rangle_{A_1} |\varphi'\rangle_{A_2} \longrightarrow
|\varphi'\rangle_{A_1} |\varphi\rangle_{A_2}.
\end{align}

When the $N$-dimensional quantum Fourier transform or the SWAP
operator are substituted for the DCNOT operator in the above three
protocols, the states of Bob's subsystem and Charlie's subsystem
after step 2 (\emph{Locking}) are also independent of $(b_1, b_2)$
and $(c_1, c_2)$. Neither Bob nor Charlie alone can learn the
encoded information from the states of their subsystems before step
4 (\emph{Unlocking}). Only after performing the inverse quantum
Fourier transform or the SWAP operator together, they can achieve
$(b_1,b_2)$ and $(c_1,c_2)$ respectively.

When the SWAP operator is used as the locking operator, Protocols 1-3 become simpler. Step 2 (\emph{Locking}) can be omitted. In step 3 (\emph{Communication}), Alice sends qu$N$it $A_1$ to Charlie
and qu$N$it $A_2$ to Bob. In step 4 (\emph{Unlocking}), Bob and
Charlie swap qu$N$it $A_1$ and qu$N$it $A_2$. However, if one of the
receivers, say Bob, is malicious, he can fool Charlie in step 4
(\emph{Unlocking}) by detaining the qu$N$it that carries $(c_1,
c_2)$ (i.e. qu$N$it $A_2$) and swapping a randomly prepared fake
qu$N$it for the qu$N$it that carries $(b_1, b_2)$ (i.e. qu$N$it
$A_1$). Then, Bob would receive $(b_1, b_2)$ from qu$N$its $A_1B$, but
Charlie would not receive achieve $(c_1, c_2)$, because qu$N$it $A_2$ is still
at Bob's side.

\section{\label{fairness} Improving the fairness of SDC}
The clients (Bob and Charlie) can decode the messages only by the
joint unlocking quantum operation, which requires either a quantum
channel, shared entanglement, or direct interaction between them. By
``direct interaction'', we mean that the clients meet and input
their particles into a physical device that can unlock the
particles, and then get their respective output particles.

In this section, we discuss the problem of achieving a global
unlocking operation between the distant parties (Bob and Charlie) by
the means of local operations and classical communication (LOCC),
some prior shared resources (such as entanglement, etc.) and/or
quantum communication. Also, we discuss the problem of the fairness
of the protocol, in case the clients do not trust each other. The
solution is a probabilistic protocol based on sequential exchange of
information (quantum, or classical with the help of prior shared
entanglement) between the clients.

\subsection{The fairness problem}
The problem of achieving a global operation by the actions of
spatially distant clients (Bob and Charlie) lies in the fact that
the unlocking operation (whether it is the inverse DCNOT, quantum
Fourier transform or SWAP\footnote{The case of the SWAP operation is
particularly interesting. Namely, after Bob and Charlie swap the
states of qubits $A_1$ and $A_2$, the very qubits do not become
entangled with each other. Nevertheless, the distant sites of Bob
and Charlie do become entangled: qubits $A_1$ are now entangled with
qubits $C$, while qubits $A_2$ are entangled with qubits $B$. The
other way to look at this is through the no-cloning theorem: there
is no way for Bob to learn the unknown state of the system $A_1B$
and transfer (swap) its partial state of $A_1$ together with its
entanglement with $B$ using only LOCC, without prior entanglement
shared with Charlie.}) has a feature that it can entangle initially
separable states. Since it is impossible to create entanglement by
the means of LOCC only, in order to implement the unlocking
operation, Bob and Charlie have to share prior entanglement, or use
a quantum channel for one of the clients to send his qu$N$it to the
other, who would then perform the unlocking operation on qu$N$its
$A_1A_2$ locally (i.e. at his site).

The problem with any such protocol is that it can be neither
simultaneous, nor symmetric, with respect to the two parties
involved, which clearly sets an unfair situation.

This is a general problem that arises within asynchronous
distributed networks~\cite{FLP85}. Namely, in order to achieve the
joint operation of the qu$N$its $A_1A_2$, clients can each perform
local measurements, and then send, conditional to the measurement
outcomes, (classical or quantum) information to the other. Since the
clients are far apart, it is impossible to achieve simultaneous
message exchange - the messages are always sent one at a time, from
one client to the other (the network is ``asynchronous''). This
means that, whatever the protocol that achieves the unlocking
operation is, there always exists the {\em last message}, say from
Bob to Charlie. But this means that, prior to sending his last
message to Charlie, Bob has all the right (quantum) information
needed to obtain his message $(b_1,b_2)$, while Charlie does not.
This is obviously unfair, with respect to Charlie.

\subsection{The solution}\label{subsec:3.b}
The situation is similar to the one presented in the contract
signing problem~\cite{EY80}. The solution, similar to the one
proposed for quantum contract signing~\cite{qcs}, is to perform the
unlocking operation on qu$N$its $A_1$ and $A_2$ (i.e., arrays of
qubits $A_{1,1}\ldots A_{1,n}$ and $A_{2,1}\ldots A_{2,n}$) in
series of steps, such that after each step, one pair of qubits are
unlocked. In the course of the unlocking stage, the clients increase
their probabilities to obtain the needed classical information
($(b_1,b_2)$ and $(c_1,c_2)$, respectively), such that at each step
one client has slightly higher probability of successful recovery
than the other. Therefore, the protocol is probabilistic, and also
fair in the sense that at each step, one client is only slightly
privileged over the other.

The information can be transferred in two equivalent ways: either by
sending quantum information (qubits) via a quantum channel, or by
teleporting qubits' quantum states using LOCC and shared
entanglement. Without the loss of generality, we will assume that
the clients are exchanging the actual qubits, rather than
teleporting their states. The only difference between the two cases
is that in the latter case, the clients use the previously shared
entanglement and exchange classical instead of quantum information.

The main problem in this approach is that a client, say Charlie,
cannot be sure if Bob sent him the right information or not.
Therefore, we introduce additional $2s$ {\em ``control qubits''}
that are used by the clients to check each other's honesty
during the unlocking stage: $s$ control qubits would be joined with
$n$ {\em ``message qubits}'' that carry the message $(b_1,b_2)$, to
form $(n+s)$ qubits $A_{1,1}\ldots A_{1,n+s}$, and given to Bob;
other $s$ control qubits joined with $n$ message qubits that carry
$(c_1,c_2)$, to form $A_{2,1}\ldots A_{2,n+s}$ and given to Charlie.

The position $Cpos_1(j) \in \{ 1,2, \ldots ,n+s \}$ of each control
qubit $j = 1,2, \ldots ,s$ within the $(n+s)$ qubits $A_{1,1}\ldots
A_{1,n+s}$ is chosen randomly by Alice and this information ($s$
integers $Cpos_1(j)$) is given to Bob; analogously, random positions
$Cpos_2(j)$ of control qubits from $A_{2,1}\ldots A_{2,n+s}$ are
given to Charlie.

The control qubits are prepared by Alice in pure states and are
uncorrelated from the message qubits. The state of each control
qubit is randomly chosen by Alice from a publicly known set
$\mathcal{S}$ of pure states. $\mathcal{S}$ is the union of two mutually
unbiased bases, computational $Z$ basis $\{ |0\rangle, |1\rangle \}$
and rotated $X$ basis $\{ |+\rangle, |-\rangle \}$, with $|\pm
\rangle = \frac{1}{\sqrt{2}}(|0\rangle \pm |1\rangle)$; therefore,
$\mathcal{S} = \{ |0\rangle, |1\rangle, |+\rangle, |-\rangle \}$.
The pure state of $j$-th control qubit in the position $Cpos_1(j)$
among qubits $A_{1,1}\ldots A_{1,n+s}$ is $Cstate_1(j)$, with
$Cstate_1(j)\in \mathcal{S}$. Analogously, the states of control
qubits from $A_{2,1}\ldots A_{2,n+s}$ are encoded by $Cstate_2(j)\in
\mathcal{S}$. Alice gives the information $\{ Cstate_1(j) | \ j =
1,2, \ldots ,s \}$ about the states of control qubits from
$A_{1,1}\ldots A_{1,n+s}$ to Bob, and the information $\{ Cstate_2(j)
| \ j = 1,2, \ldots ,s \}$ about the states of control qubits from
$A_{2,1}\ldots A_{2,n+s}$ to Charlie.

Therefore, prior to Alice sending the message to clients (via
sending the qubits $A_{1,1}\ldots A_{1,n+s}$ to Bob and
$A_{2,1}\ldots A_{2,n+s}$ to Charlie), Bob has, apart from qu$N$it
$B$, the information $\{ Cpos_1(j) | \ j = 1,2, \ldots ,s \}$ of the
positions and $\{ Cstate_1(j) | \ j = 1,2, \ldots ,s \}$ of the the
states of the control qubits from $A_{1,1}\ldots A_{1,n+s}$, and
analogously for Charlie.

During the initialisation phase, Alice, Bob and Charlie share two pairs
of entangled qu$N$its, denoted as
\begin{align}
|\phi(00)\rangle_{A_1B}\otimes|\phi(00)\rangle_{A_2C}.
\end{align}
Alice and Bob also share classical information of the control
qubits: $\{ Cpos_1(j) | \ j = 1,2, \ldots ,s \}$ and $\{ Cstate_1(j) |
\ j = 1,2, \ldots ,s \}$. Alice and Charlie also share classical
information $\{ Cpos_2(j) | \ j = 1,2, \ldots ,s \}$ and $\{
Cstate_2(j) | \ j = 1,2, \ldots ,s \}$.

The protocol of simultaneous dense coding of classical messages
$(b_1,b_2)$ and $(c_1,c_2)$ works as follows:

(1) \emph{Encoding.} Alice performs unitary operators $U(b_1b_2)$ on
qu$N$it $A_1$ and $U(c_1c_2)$ on qu$N$it $A_2$ to encode $(b_1,b_2)$
and $(c_1,c_2)$, respectively.

(2) \emph{Locking.} Alice joins the $s$ control qubits with $n$
message qubits $A_{1,1}\ldots A_{1,n}$, to form the ordered set of
$(n+s)$ qubits $A_{1,1}\ldots A_{1,n+s}$. The $j$-th control qubit
is prepared in $Cstate_1(j)\in \mathcal{S}$ and is in the position
$Cpos_1(j)$, while the relative positions of the $n$ message qubits
are the same as before. Analogously, she forms the set
$A_{2,1}\ldots A_{2,n+s}$.

For $j=1$ to $n+s$, Alice applies the locking operator on
$A_{1,j}A_{2,j}$.

(3) \emph{Communication.} Alice sends $(n+s)$ qubits $A_{1,1}\ldots
A_{1,n+s}$ to Bob, and $(n+s)$ qubits $A_{2,1}\ldots A_{2,n+s}$ to
Charlie.

(4) \emph{Unlocking.} For $j=1$ to $n+s$, Bob sends qubit $A_{1,j}$
to Charlie, and then Charlie returns $A_{1,j}$ to Bob after
performing the unlocking operator on $A_{1,j}A_{2,j}$ at his site.

If $\exists k, Cpos_1(k)=j$ (i.e., if the $j$-th qubit given to Bob is a controlled one), Bob measures $A_{1,j}$ in either $X$ or
$Z$ basis, according to $Cstate_1(k)$. If the measurement result does
not match $Cstate_1(k)$, he knows that Charlie did not return the
real $A_{1,j}$ and stops the unlocking stage.

Analogously, if $\exists k, Cpos_2(k)=j$, Charlie measures $A_{2,j}$
in either $X$ or $Z$ basis, according to $Cstate_2(k)$. If the
measurement result does not match $Cstate_2(k)$, he knows that Bob
did not send the real $A_{1,j}$ and stops the unlocking stage.

The remaining $n$ qubits received at Bob's site form the ordered
set $A_{1,1}\ldots A_{1,n}$, and the remaining $n$ qubits received
at Charlie's site form the ordered set $A_{2,1}\ldots A_{2,n}$.

(5) \emph{Decoding.} Bob and Charlie measure qu$N$its $A_1B$ and
qu$N$its $A_2C$ in the Bell basis respectively to achieve
$(b_1,b_2)$ and $(c_1,c_2)$.

The unlocking of qubits $A_{1,1}\ldots A_{1,n+s}$ and $A_{2,1}\ldots
A_{2,n+s}$ is done in $(n+s)$ steps, such that in each step one pair
of qubits is unlocked. The order of qubits must be maintained.
Without the loss of generality, we assume that the unlocking
operation is done at Charlie's site. We will assume that Bob is an
honest client who sends qubits $A_{1,1}\ldots A_{1,n+s}$ to Charlie,
as long as he is convinced that Charlie is returning the exact
resulting qubit after the local unlocking operation at Charlie's
site. The way to check if Charlie is indeed doing so is the
following: in each step $Cpos_1(j)$ of the unlocking stage (i.e., when the qubit given by Alice to Bob is a controlled one), Bob
measures the state of the qubit received back from Charlie during that step in one of the
two mutually unbiased bases -- $Z$ if $Cstate_1(j)$ is in $\{
|0\rangle, |1\rangle \}$, $X$ otherwise. If the measurement result
matches the classical information $Cstate_1(j)$, i.e. if Bob's
outcome is consistent with Charlie returning the control qubit in
$Cstate_1(j)$, Bob continues with the unlocking stage. Otherwise, it
means that Charlie did not return the control qubit in $Cstate_1(j)$
and Bob stops the unlocking stage. Since Charlie does not know the
positions of control qubits from $A_{1,1}\ldots A_{1,n+s}$, he has
to return all of the resulting qubits to Bob. Otherwise, he will
inevitably return some qubits as control ones in states different
from those prepared by Alice, and Bob will, with high probability,
be able to detect it.

Note that the whole analysis is done for the ideal case where no
measurement errors or decoherence effects occur. The existence of
measurement errors will set the threshold value $\eta > 0$ for the
allowed number of wrong results obtained on control qubits (results
inconsistent with Charlie sending the control qubits in
$Cstate_1(j)$), which will increase the number $s$ of the control
qubits. After receiving, in the course of exchange, $k \leq s$
controlled qubits, we require that not more than $\eta k$ wrong
results are obtained. The parameter $\eta$ is determined by the
experimental set-up, which sets the probability of obtaining the
wrong result when the right qubit is sent.

The protocol is {\em optimistic}~\cite{ASW97}: if both clients are
honest, if they execute the protocol by unlocking the qubits Alice
gave them, upon finishing the unlocking stage both parties will have
the whole information sent by Alice.

Unfortunately, if we insist on the {\em perfect} fidelity of data
transmission, situations when one client (say Charlie) knows that he
obtained the whole information, while the other hasn't, clearly puts
Bob in a disadvantaged situation. For example, if the unlocking
operation is done at Charlie's site, Charlie can decide not to
return the last resulting qubit to Bob, which would leave him
without the whole state $|\phi(b_1b_2)\rangle_{A_1B}$, and thus
$(b_1,b_2)$, in case $Cpos_1(s)< n+s$. The perfect fidelity was for
the same reasons relaxed for quantum contract signing as well, by
introducing the factor $\alpha < 1$ of the required fraction of
correct results for the total number $N$ of qubits sent to a client~\cite{qcs}. Therefore, we will also introduce factor $\alpha < 1$
and require that a client has correct values of $\alpha (2n)$
message bits sent to him by Alice.

The above protocol is clearly {\em probabilistic}. At each step of
the unlocking stage, Bob has a finite probability, which increases
with the execution of the protocol, to obtain correct values of
$2\alpha n$ bits of the message $(b_1,b_2)$ that Alice sent him, and
analogously for Charlie.

Due to its probabilistic nature, during the execution of the
protocol one client is always privileged over the other. By
privileged, we mean that one client has higher probability of
obtaining the required fraction $\alpha$ of the message Alice
encoded for him, than the other. The difference between the clients'
probabilities to obtain the information sent by Alice is due to the
random distribution of control qubits. On the other side, since
neither of the clients know the distribution of both sets of control
qubits, even though one of them might be privileged over the other
at a certain step of the unlocking stage, he would not know that:
the protocol is {\em a priori} symmetric with respect to the
clients.

Yet, the protocol is {\em fair}~\cite{ben:90}: throughout the
execution, one client is only slightly privileged over the other,
the difference being smaller with the increase of $s$ and could be
made arbitrarily small for big enough $s$. The argument here is
exactly the same as the one presented for the fairness of the
quantum contract signing protocol~\cite{qcs}.

Namely, the role of the control qubits is to signal possible {\em
cheating} of a client. By cheating, we mean not sending the qubits
received by Alice. For example, if one of the clients, say Charlie,
starts sending qubits each randomly in one of four states $\{
|0\rangle, |1\rangle, |+\rangle, |-\rangle \}$, he will send wrong
message qubits $A_1$ sent by Alice, but also wrong control qubits.
Therefore, for each control qubit he will have a finite probability
of $1/2$ of being detected cheating (he did not send the right
control qubit). Therefore, since Bob's probability $p_w$ to detect a
wrong qubit (Charlie's cheating) is approaching to one exponentially fast, $p_w =
1 - (1/2)^m$, where $m$ is a number of controlled qubits that are
sent as random, it is not difficult to estimate the expected
difference between Bob and Charlie, depending on the total number
$n$ of the message qubits to be sent from one client to another and the
number $s$ of the controlled qubits given to each client.

One simply has to estimate the expected value for $m$, the expected
number $\langle m \rangle$ of wrongly sent control qubits after
which Bob notices the cheating and stops communication. During that
time, Bob received certain number $w$ of random qubits as the
message qubits, while Charlie was still receiving the proper ones.
Therefore, he is in advantage by having about $w/2$ more correct
values for message bits (even when guessing bit values at random,
one has $1/2$ of the probability to be correct). The value $w$ is
a simple function of $m$, $w=w(m)$, and the dependence on $m$ can be
straightforwardly determined by $n$ and $s$. For big enough $n$,
$w(\langle m \rangle) << n$ and therefore Charlie's advantage is
negligible.

Of course, Charlie can try to send qubits in some other states
(completely random, etc.), but as long as he's not sending the right
qubits in the right states, there will be a finite probability $p>0$
that Bob detects cheating in a single-shot measurement, so that his
probability to detect cheating will again exponentially fast
approach to one in the number of wrongly sent control qubits, making
Charlie's privilege arbitrarily small, for big enough $n$. Charlie
can try to decrease the value of $p$, but can never make it zero.
Otherwise, he would be able to perfectly distinguish between the
non-orthogonal states and this would violate the security of the
BB84 cryptographic protocol~\cite{bb84}, for example.

In the case of quantum contract signing~\cite{qcs} the role of message and
control qubits was given to the same $N$ qubits sent by a trusted
party (in our case Alice) to clients (in our case Bob and Charlie),
while in the case of simultaneous dense coding the two roles are
given to separate sets of qubits. This will introduce a slight
change in the expressions for the probabilities involved in
calculation, but this change is minor, conceptually straightforward
to calculate and does not affect the main result of protocol's
fairness. Nevertheless, knowing the exact expressions for the
probabilities in case of different cheating strategies is of crucial
importance, and may technically be quite non-trivial, which clearly
presents interesting and challenging topics for future research.

The fairness condition can be even straightened, as was done in~\cite{qcs}, by requiring the negligible {\em probability to cheat}:
the probability that one client has $2\alpha n$ right bits, while
the other does not. This quantity may be quite relevant in various
possible scenarios, like in the case of signing contracts for buying
and selling the goods on the market (see~\cite{qcs}). While for a
fixed value of $\alpha$ the probability to cheat may be as high as
$1/4$, if it is unknown to the clients and chosen randomly by Alice
from a certain interval $I_{\alpha} \subset (1/2,1)$, the
probability to cheat can be made as small as needed, for big enough
$n$ (see~\cite{qcs}).

The locking and unlocking operators performed on two qu$N$its (i.e.,
two arrays of qubits) in this section are actually products of
two-qubit locking and unlocking operators. The $N$-dimensional DCNOT
operator and the SWAP operator are of this kind. But the
$N$-dimensional quantum Fourier transform cannot be done
qubit(pair)-by-qubit(pair). Therefore, the quantum Fourier transform cannot
be used to implement the SDC protocol proposed in this section.

\section{\label{security} Improving the security of SDC}
We now consider the security of simultaneous dense coding. We assume
that in the initialisation phase, the entangled pairs have been
securely distributed among Alice, Bob and Charlie. Because only part
of the entangled pairs travel through the quantum channel, an
outsider knows nothing about the encoded information. If the two
receivers are honest and follow the protocols exactly, they must
collaborate to achieve their respective information. However, if one
of the receivers, say Bob, is dishonest and has the ability to
intercept and resend the qubits going through the quantum channel
between Alice and Charlie, he can obtain $(b_1,b_2)$ without
collaborating with Charlie by the following intercept-resend attack:
(1) intercept qu$N$it $A_2$; (2) perform the unlocking operation on
qu$N$its $A_1A_2$; (3) measure qu$N$its $A_1B$ to obtain
$(b_1,b_2)$; (4) perform the locking operation on qu$N$its $A_1A_2$;
(5) send qu$N$it $A_2$ back to Charlie.

To detect such a cheating behaviour, we insert additional $2r$
``\emph{detect qubits}'' into the array of $2n$ message qubits
during the communication phase: $r$ detect qubits would be joined
with $n$ message qubits that carry the message $(b_1,b_2)$, to form
$(n+r)$ qubits $A_{1,1}\ldots A_{1,n+r}$, and given to Bob; other
$r$ detect qubits joined with $n$ message qubits that carry
$(c_1,c_2)$, to form $A_{2,1}\ldots A_{2,n+r}$ and given to Charlie.

The position $Dpos_1(j) \in \{ 1,2, \ldots ,n+r \}$ of each detect
qubit $j = 1,2, \ldots ,r$ within the $(n+r)$ qubits $A_{1,1}\ldots
A_{1,n+r}$ is chosen randomly by Alice, and analogously for
$Dpos_2(j)$ of each detect qubit within $A_{2,1}\ldots A_{2,n+r}$.

The state of each detect qubit is randomly chosen by Alice from set
$\mathcal{S} = Z \cup  X = \{ |0\rangle, |1\rangle, |+\rangle, |-\rangle \}$.
The pure state of $j$-th detect qubit in
the position $Dpos_1(j)$ among qubits $A_{1,1}\ldots A_{1,n+r}$ is
$Dstate_1(j)$, with $Dstate_1(j)\in \mathcal{S}$. Analogously, the
states of detect qubits from $A_{2,1}\ldots A_{2,n+r}$ are encoded
by $Dstate_2(j)\in \mathcal{S}$.

Only after the transmission of all the message qubits and detect
qubits, Alice tells Bob and Charlie the positions and the bases of
the detect qubits and requires them to return the results of the measurements performed on the detect qubits. Therefore Alice can check if after the transmission the
states of the detect qubits have been altered by a dishonest client.

Because the position of the detect qubits in the array are chosen
randomly by Alice, curious Bob does not know which qubits are the
message qubits. If his intercept-resend attack involves a detect
qubit, the state of the detect qubit may probably be changed, the
probability of being detected grows exponentially with the increase
of $r$ and could be made arbitrarily large for big enough $r$.

During the initialisation phase, Alice, Bob and Charlie share two pairs
of entangled qu$N$its, denoted as
\begin{align}
|\phi(00)\rangle_{A_{1}B}\otimes|\phi(00)\rangle_{A_{2}C}.
\end{align}

The protocol of simultaneous dense coding of classical messages
$(b_1,b_2)$ and $(c_1,c_2)$ works similarly as in the previous case. For completeness, in~\ref{combination} we present a detailed description of the combination of the two strategies that assure fair and secure simultaneous dense coding protocol.

We summarise the above protocols in Table 1 in order to
demonstrate their qualitative and quantitative differences.
The original SDC protocol in~\cite{SQ10} and the $N$-dimensional SDC protocol in Sec. \ref{n-dimension} are vulnerable to intercept-resend attacks and require the receivers to perform the unlocking operation at the same site, otherwise the fairness problem arises.
The improved SDC protocol in Sec. \ref{fairness} can guarantee fairness in the unlocking phase even if the receivers are separated spatially, by implementing the unlocking operation in series of steps of communication and introducing additional control qubits for cheat detection.
On the other hand, the improved SDC protocol in Sec. \ref{security} can guarantee security against the intercept-resend attack in the communication phase, by introducing additional detect qubits in the communication phase.
The protocol in Appendix incorporates both fairness and security properties.
We can also see from the table that the cost of the protocol increase with these improvements.


\begin{table}
\begin{center}
\caption{Comparison between different protocols}
\label{tab1}
\begin{tabular}{ccc}
\hline
Protocol & Properties & Extra cost\\
\hline
protocol in Sec. \ref{n-dimension} and Ref. \cite{SQ10} & & \\
protocol in Sec. \ref{fairness} & fair & $s$ control qubits\\
protocol in Sec. \ref{security} & secure & $r$ detect qubits\\
protocol in Appendix & fair, secure & $s$ control qubits  $+r$ detect qubits\\
\hline
\end{tabular}
\end{center}
\end{table}

\section{\label{application} Applications to contract signing}

Contract signing \cite{EY80} is a security task involving two
parties, Charlie and Bob, that do not trust each other and want to
exchange a common contract signed with each other's signature. At
the end of the protocol Charlie should have the contract signed by
Bob and vice-versa. The purpose of the signed contract is to
bind the parties to the terms of the contract, which can be
enforced by a judge (Alice). The real challenge to this problem is
when both Charlie and Bob are physically apart and want to remotely sign the contract. This situation is becoming more and more
common due to e-business, and may lead to fraud. For instance, if Bob
gets the contract signed by Charlie without committing himself,
Charlie and Bob are in an unfair situation. By having Charlie's
commitment, Bob is able to appeal to Alice to enforce the contract,
while Charlie has no means to do the same, since he does not possess
the contract signed by Bob. Note that even if Bob did not commit,
but having Charlie's commitment, puts him in a position to later in
time choose whether to bind the contract or not, while Charlie has
no power to do either of the two.

A simple solution to this unfair situation is to have a trusted
third party (again Alice) mediating the transaction -- Bob sends to
Alice the contract signed by him and Charlie does the same; then
Alice exchanges the contracts only after she has received both of
the commitments. Note that this procedure increases significantly
the cost of remote contract signing, as Alice's time and resources
are expensive. What is particularly costly is Alice being constantly
online and alert waiting for the clients to contact her. Also,
avoiding the communication with the trusted party at the very moment
of determining a contract and committing to it removes the danger of
overloading Alice and creating a bottleneck.

Unfortunately, it has been shown that the attendance of Alice is
mandatory \cite{EY80,FLP85}, if the protocol is to fulfil the
following two important properties:
\begin{itemize}
    \item {\em fairness}: either both parties get each other'
    commitment or none gets;
    \item {\em viability}: if both
    parties behave honestly, they will both get each other's
    commitments.
\end{itemize}

One way to overcome this difficulty is to consider {\em optimistic
protocols} that do not require communication with Alice unless
something wrong comes up \cite{ASW97}. Another workaround is to
relax the fairness condition, allowing one agent to have $\epsilon$
more probability of binding the contract than the other agent
({\em probabilistic fairness}). In this case, for an arbitrary small
$\epsilon$ solutions have been found where the number of exchanged
messages between the agents is minimised \cite{ben:90}.

In this section we present a probabilistically fair quantum protocol based on  a secure SDC for remote agents, such as the one described in~\ref{combination} (a combination of protocols presented in Section~\ref{subsec:3.b} and Section~\ref{security}). In such protocol,  Alice, the trusted party, does not interact with the signing parties while they (Bob
and Charlie) are determining the contract and then committing to it
(the Exchange Phase). Note that in the below construction of a contract signing scheme, a SDC protocol is used iteratively many times as a {\em black box}. Thus,  {\em any} protocol that achieves simultaneous transmission of (densly) coded information that satisfies the fairness and the security conditions, as presented in Section~\ref{fairness} and Section~\ref{security}, respectively, allows for constructing probabilistically fair contract signing scheme.

In order to lower the use of resources, it is possible to map long
contracts into messages of a small fixed size, say of $k$ bits. Such
short messages (digests) are obtained by so called hash functions
(such as SHA1) and are well established in the field of cryptography~\cite{hac}. Hash functions are not injective, that is, there exist
pairs of different messages $x,x'$ with the same digest $d$.
Nevertheless,  given a message $x$ and its digest $d$, it is
computationally hard to find a different message $x'\neq x$ with the
same digest $d$. For this reason, digests can be used to identify a
message. From this point on, instead of contracts themselves we
consider their digests, obtained by some hash function, with $k$
bits, say $(b_1\dots b_k)$, where $b_i\in\{0,1\}$.

We also assume that
Alice can sign messages using some public key signing scheme (such
as DSS, for more detail see \cite{hac}). In short, a public key
signature scheme for Alice is a pair of functions $(sig,ver)$
together with a pair of keys $(A,\hat{A})$ where $A$ is the private
key of Alice (known only by Alice) and $\hat{A}$ is the
corresponding public key (known by Alice, Bob and Charlie). For
Alice to sign a message $m$, she uses her private key $A$ and obtains
the signature $sig_A(m)$. Bob (or Charlie) verifies if $sig_A(m)$ is
the message $m$ signed by Alice by checking whether
$ver_{\hat{A}}(m,sig_A(m))=1$. If $ver_{\hat{A}}(m,sig_A(m))\neq 1$
then $sig_A(m)$ does not correspond to the signature of Alice over
$m$. Signing can be seen as an encryption, unique to Alice: only she
can do it using her private key $A$. But Bob and Charlie can, given
the message $m$ and a signature $s$, by using Alice's public key
$\hat{A}$ verify whether $s$ is indeed the signature of $m$ by Alice or
not.

In our contract signing protocol, Alice does not know a priori to
which particular contract/digest $(b_1\dots b_k)$ Bob and Charlie
are going to agree upon. However, at the end of the protocol Alice
(and also Bob and Charlie) needs some irrefutable proof of the
particular contract that was agreed upon. Moreover, we do not want
Alice to be contacted during the exchange phase. A solution for this
problem is for Alice to produce $4k$ triples
$\{(b,i,\textrm{Bob}),(b,i,\textrm{Charlie}) : b\in\{0,1\},i=1\dots
k\}$ that can be used to represent any particular contract
$(b_1\dots b_k)$ that Bob and Charlie will agree upon latter. Next,
Alice signs those 4$k$ triples and prepares 2$k$ SDC protocols for
Bob and Charlie, where in each of these SDC protocols the messages
to be simultaneously received are $sig_{A}(b,i,\textrm{Charlie})$ by
Bob and $sig_{A}(b,i,\textrm{Bob})$ by Charlie, with $b\in\{0,1\}$
and $i\in \{1\dots k\}$.

Thus, Bob can enforce the contract $(b_1\dots b_k)$ against Charlie (and no other person) if he shows to Alice the signatures $sig_{A}(b_1,1,\textrm{Charlie})\dots sig_{A}(b_k,k,\textrm{Charlie})$. By using SDC, it is possible to force Bob to obtain this information (and no other) only with the collaboration of Charlie, and vice-versa. In detail the contract signing protocol based on SDC works as follows:

\bigskip

\noindent
{\em Initialization Phase:}\ \\[2mm]
\noindent
1. Alice signs the  following $4k$ messages: $(0,i,\textrm{Bob})$, $(1,i,\textrm{Bob})$, $(0,i,\textrm{Charlie})$, $(1,i,\textrm{Charlie})$ with $i=1,2,\dots ,k$.\\[2mm]
\noindent 2. Alice arranges for $2k$ different SDC's for the case of
distant parties such that the message to be sent to Bob in one of
such SDC's is $sig_A(b,i,\textrm{Charlie})$ and to Charlie is
$sig_A(b,i,\textrm{Bob})$ for $i=1,2,\dots ,k$ and $b=0,1$.\\[4mm]
\noindent{\em Exchange Phase:}\ \\[2mm]
1. Bob and Charlie agree on contract $(b_1\dots b_k)\in\{0,1\}^k$.\\[2mm]
2. For $i=1$ to $k$, Bob and Charlie collaborate to obtain from the
entangled quNits of the $2k$ SDC's the messages $sig_A(b_i,i,\textrm{Charlie})$ and $sig_A(b_i,i,\textrm{Bob})$,
respectively,
and ignore the remaining quantum data
sent by Alice. Thus, at the end of the SDC's Bob has
$sig_{A}(b_i,i,\textrm{Charlie})$, for $i=1\dots k$, and mutatis
mutandis for Charlie.
\\[4mm]
\noindent{\em Binding Phase:}\ \\[2mm]
1. Alice enforces contract $(b_1\dots b_k)$ when either Bob presents
$sig_A(b_1,1,\textrm{Charlie})\dots sig_A(b_k,k,\textrm{Charlie})$
or Charlie presents
$sig_A(b_1,1,\textrm{Bob})\dots sig_A(b_k,k,\textrm{Bob})$.

\bigskip

The size of each signature, say $sig_A(b_i,i,\textrm{Charlie})$, is
given by the length of a message a client receives in each SDC,
which is $2n$ bits (the factor 2 comes from the fact that the coding
is dense). Since each SDC is probabilistic, a client only needs to
present to Alice $\alpha_i (2n)$ bits,  with
$\frac{1}{2}<\alpha_i<1$, of the signature
$sig_A(b_i,i,\textrm{Charlie})$ for each $i=1\dots k$. Moreover, if
each $\alpha_i$ is random we achieve even stronger fairness
condition, namely the {\em expected} probability to cheat on each SDC can
be made arbitrarily small (the probability to cheat is the
probability that an agent obtains at least $\alpha_i(2n)$ bits of a
signature and the other does not \cite{qcs}).

The contract signing protocol described above, unlike the one
presented in \cite{qcs}, allows for Bob and Charlie to determine the
contract after the Initialisation Phase. This is due to the fact
that public key signatures were introduced. However, this
introduction leads to a poorer security assumption, as public key
signatures are only computationally secure (as well as hash
functions).

We can improve the security of the above protocol by removing both
hash functions and public key signatures. Removing hash function
accounts to consider a large enough $k$ such that  all potential
contracts by Bob and Charlie would fit in $k$ bits. To remove public
key signatures, which are not perfectly secure, Alice can, during
the initialisation phase, share a symmetric key $k_{AB}$ with Bob
and another, $k_{AC}$, with Charlie. These symmetric keys might be
made perfectly secure by using one-time pad cryptosystems (see, for
example \cite{hac}). Then, in each SDC the message that Bob receives
is $k_{AB}(b_i,i,r_i(b_i))$ and the message that Charlie receives is
$k_{AC}(b_i,i,r_i(b_i))$. Here, $k_{AB}(m)$ is the encryption of $m$
with the symmetric key $k_{AB}$ and $r_i(b_i)$ is a random string
for each $i$ and $b_i$, sampled and known only by Alice, that is
associated to a contract between Bob and Charlie. So, for Bob to
enforce contract $(b_1\dots b_k)$ against Charlie, he has to present
to Alice the random numbers $r_1(b_1)\dots r_k(b_k)$. Note that in
this case Alice has to store all these random numbers $r_i(b_i)$ in
her private memory keeping in mind to whom they are associated. In
this way the protocol's perfect security is obtained by the laws of
physics, which is stronger than computational security used in
classical protocols.

\section{\label{teleportation} $N$-dimensional simultaneous teleportation}

Dense coding and quantum teleportation are dual protocols intimately linked to each other, both in their purpose as well as in the construction. The former is used to transmit (densely coded) classical information, while the latter transfers quantum states (i.e., quantum information). In Introduction, as a motivation for SDC the following situation was described: an agent, say a boss of a company, needs for two employees to simultaneously carry out two different confidential tasks, such that each employee is unaware of the other's activity. SDC allows that the two tasks, encoded by classical information, are confidentially communicated to the employees. Similarly, one could imagine a situation in which the employees are requested to each execute a predetermined quantum protocol (say, a quantum computation algorithm) for a given confidential initial state. The solution for such situation is achieved by a simultaneous teleportation protocol.

Simultaneous teleportation was proposed by Wang {\it et al}~\cite{WYGL08}, in which
all the receivers simultaneously obtain their respective quantum
states from the sender. In their scheme, the sender first performs a
locking operation to entangle the particles from two independent
quantum entanglement channels, and therefore the receivers cannot
restore their quantum states separately before performing the
unlocking operation together. The locking operator is composed of
the Hadamard and the CNOT operators.

Ref.~\cite{SQ10} showed that the quantum Fourier transform can
alternatively be used as the locking operator in simultaneous
teleportation. In this section, we further investigate simultaneous
teleportation of qu$N$its using the $N$-dimensional quantum Fourier
transform.

In the task of $N$-dimensional simultaneous teleportation, Alice
intends to teleport the unknown qu$N$its
$|\varphi_t\rangle_{T_t}=\sum_{s=0}^{N-1}\alpha_{t,s}|s\rangle_{T_t}$
to Bob$_t\ (1\leq t\leq M)$ under the condition that all the
receivers must collaborate to simultaneously obtain
$|\varphi_t\rangle_{T_t}$.

In the initialisation phase, Alice shares with each Bob$_t$ a pair of
entangled qu$N$its $|\phi(00)\rangle_{A_tB_t}$, where subscript
$A_t$ denotes Alice's qu$N$it, and subscript $B_t$ denotes Bob$_t$'s
qu$N$it. The initial quantum state of the composite system is
\begin{align}
|\chi (0)\rangle & =  \bigotimes_{t=1}^{M} |\phi(00)\rangle_{A_tB_t}
\bigotimes_{t=1}^{M} |\varphi_t\rangle_{T_t} \nonumber\\ & 
= \frac{1}{\sqrt{N^M}} \sum_{j=0}^{N^M-1} |j\rangle_{A_1\ldots
A_M} |j\rangle_{B_1\ldots B_M} \bigotimes_{t=1}^{M}
|\varphi_t\rangle_{T_t},
\end{align}
where $j$ is a base-$N$-number which can be written as $j_1j_2\ldots
j_M, j_t\in\{0,1,\ldots, N-1\}.$

The protocol for simultaneous teleportation of qu$N$its consists of
five steps:

(1) {\it Locking.} Alice performs the $N$-dimensional quantum
Fourier transform
\begin{align}
|j\rangle_{A_1\ldots A_M}\longrightarrow \frac{1}{\sqrt{N^M}}
\sum_{k=0}^{N^M-1} e^{\frac{2\pi i}{N^M}jk} |k\rangle_{A_1\ldots
A_M}
\end{align}
on qu$N$its $A_1\dots A_M$ to lock the entanglement channels. After
that, the state of the composite system becomes
\begin{align}
|\chi(1)\rangle = & \frac{1}{\sqrt{N^M}} \sum_{j=0}^{N^M-1}
QFT_{A_1\ldots A_M}|j\rangle_{A_1\ldots A_M} |j\rangle_{B_1\ldots
B_M} \bigotimes_{t=1}^{M} |\varphi_t\rangle_{T_t}\nonumber\\
= & \frac{1}{\sqrt{N^M}} \sum_{j=0}^{N^M-1} \frac{1}{\sqrt{N^M}}
\sum_{k=0}^{N^M-1} e^{\frac{2\pi i}{N^M}jk} |k\rangle_{A_1\ldots
A_M}  |j\rangle_{B_1\ldots B_M} \bigotimes_{t=1}^{M}
|\varphi_t\rangle_{T_t}.
\end{align}

(2) {\it Measuring.} Alice measures each pair of qu$N$its $A_tT_t$
in the $N$-dimensional Bell basis.
\begin{align}
  &  \bigotimes_{t=1}^{M}\ _{A_tT_t} \langle \phi
(x_ty_t)|\chi(1)\rangle\nonumber\\
  = &  \frac{1}{N^M} \Bigg( \bigotimes_{t=1}^{M} \sum_{j=0}^{N-1}
e^{-\frac{2\pi i}{N}jx_t}\ _{A_t}\langle j + y_t|_{T_t}\langle
j|\Bigg) \nonumber\\
  &   \Bigg[ \sum_{k=0}^{N^M-1} \Big(
\bigotimes_{t=1}^{M}|k_t\rangle_{A_t}|\varphi_t\rangle_{T_t}\Big)
\frac{1}{\sqrt{N^M}}\sum_{j=0}^{N^M-1}e^{\frac{2\pi
i}{N^M}jk}|j\rangle_{B_1\ldots
B_M}\Bigg] \nonumber\\
  = &  \frac{1}{N^M} \sum_{k=0}^{N^M-1}\Bigg(
\prod_{t=1}^{M}\sum_{j=0}^{N-1}e^{-\frac{2\pi i}{N}jx_t} \langle j +
y_t|k_t\rangle \langle j| \sum_{s=0}^{N-1}
\alpha_{t,s}|s\rangle\Bigg)   QFT_{B_1\ldots B_M}
|k\rangle_{B_1\ldots B_M}\nonumber\\
  = &  \frac{1}{N^M} QFT_{B_1\ldots B_M} \sum_{k=0}^{N^M-1}
\bigotimes_{t=1}^{M} e^{-\frac{2\pi i}{N}x_t(k_t- y_t)}
\alpha_{t,k_t - y_t} |k_t\rangle_{B_t}\nonumber\\
  = &  \frac{1}{N^M} QFT_{B_1\ldots B_M} \sum_{k=0}^{N^M-1}
\bigotimes_{t=1}^{M} e^{-\frac{2\pi i}{N}x_tk_t}
\alpha_{t,k_t} |k_t + y_t\rangle_{B_t}\nonumber\\
  = &  \frac{1}{N^M} QFT_{B_1\ldots B_M} \sum_{k=0}^{N^M-1}
\bigotimes_{t=1}^{M} \alpha_{t,k_t} X^{y_t}(Z^\dagger)^{x_t}|k_t\rangle_{B_t}\nonumber\\
  = & \frac{1}{N^M} QFT_{B_1\ldots B_M} \bigotimes_{t=1}^{M}
X^{y_t}(Z^\dagger)^{x_t} \sum_{s=0}^{N-1} \alpha_{t,s}
|s\rangle_{B_t},
\end{align}
where $X: |j\rangle \longrightarrow |j + 1\rangle$ is the shift
operator, $Z: |j\rangle \longrightarrow e^{\frac{2\pi
i}{N}j}|j\rangle$ is the rotation operator.

If the measurement result of qu$N$its $A_tT_t$ is
$|\phi(x_ty_t)\rangle$, the state of qu$N$its $B_1\dots B_M$
collapses into
\begin{align}
|\chi(2)\rangle = QFT_{B_1\ldots B_M} \bigotimes_{t=1}^{M}
X^{y_t}(Z^\dagger)^{x_t} |\varphi_t\rangle_{B_t}.
\end{align}

(3) {\it Communication.} Alice sends the measurement result $(x_t,
y_t)$ to each Bob$_t$.

(4) {\it Unlocking.} All the receivers collaborate to perform the
inverse quantum Fourier transform on qu$N$its $B_1\dots B_M$, and
the state of qu$N$its $B_1\dots B_M$ becomes

\begin{align}
|\chi(3)\rangle = QFT_{B_1\ldots B_M}^\dagger |\chi(2)\rangle
 = \bigotimes_{t=1}^{M} X^{y_t}(Z^\dagger)^{x_t}
|\varphi_t\rangle_{B_t}.
\end{align}

(5) {\it Recovering.} Each Bob$_t$ performs
$Z^{x_t}(X^\dagger)^{y_t}$ on qu$N$it $B_t$ to obtain
$|\varphi_t\rangle$.

\section{\label{conclusion} Conclusions}
Dense coding~\cite{BW92} and teleportation~\cite{BBCJPW93} are important quantum communication tasks. Simultaneous dense coding~\cite{SQ10} and simultaneous teleportation~\cite{WYGL08}, which guarantee that the receivers simultaneously achieve their respective
information from one sender, may be relevant and useful for
improvement of some models or tasks of quantum communication. In
this paper, we have given a number of new results on simultaneous
dense coding and teleportation. More specifically, we have given
three protocols for simultaneous dense coding utilising different
$N$-dimensional quantum states (i.e., Bell state, W state, and GHZ
state). Besides the quantum Fourier transform, we have introduced
two new locking operators (i.e. the double controlled-NOT operator
and the SWAP operator) for simultaneous dense coding. Then we have
analysed the fairness and the security of the simultaneous dense
coding protocol and proposed a protocol which guarantees both the fairness
and the security, thus allowing for mutually distant receivers to execute the protocol. We have shown that any fair simultaneous dense coding scheme that is secure against intercept-resend attack can be used to implement a fair contract signing protocol. In addition, we have shown that the $N$-dimensional quantum Fourier transform can act as the locking operator in simultaneous teleportation of qu$N$its.

\section*{Acknowledgments}
The authors are grateful to the anonymous referee for invaluable comments and suggestions that help us improve the quality of the paper.
This work is supported in part by the National Natural Science Foundation (Nos. 61272058, 61572532, 61502179), the Natural Science Foundation of Guangdong Province of China (No. 10251027501000004, 2014A030310265), the Research Foundation for the Doctoral Program of Higher School of Ministry of Education of China (No. 20100171110042),
FCT project UID/EEA/50008/2013 and IT initiatives PQDR (Probabilistic, Quantum and Differential Reasoning) and CaPri (Capacity and Privacy with Quantum Continuous Variables).


\appendix

\section{\label{combination} Combination of the two strategies}
In this appendix, we combine in one protocol the two strategies described in the
above two sections: (i) if Bob and Charlie have the
ability to intercept and resend the qubits going through the quantum
channel, Alice can detect such behaviour and interrupt the protocol; (ii) if Bob and Charlie are spatially separated, they can fairly decode their
respective messages simultaneously through communication.

In the initialisation phase, Alice, Bob and Charlie share two pair
of entangled qu$N$its, denoted as
\begin{align}
|\phi(00)\rangle_{A_{1}B}\otimes|\phi(00)\rangle_{A_{2}C}.
\end{align}
Alice and Bob also share classical information of the control
qubits: $\{ Cpos_1(j) | \ j = 1,2, \ldots ,s \}$ and $\{
Cstate_1(j) | \ j = 1,2, \ldots ,s \}$. Alice and Charlie also
share classical information $\{ Cpos_2(j) | \ j = 1,2, \ldots ,s
\}$ and $\{ Cstae_2(j) | \ j = 1,2, \ldots ,s \}$.

The protocol of simultaneous dense coding of classical messages
$(b_1,b_2)$ and $(c_1,c_2)$ works as follows:

(1) \emph{Encoding.} Alice performs unitary operators $U(b_1b_2)$ on
qu$N$it $A_1$ and $U(c_1c_2)$ on qu$N$it $A_2$ to encode $(b_1,b_2)$
and $(c_1,c_2)$, respectively.

(2) \emph{Locking.} Alice joins the $s$ control qubits with $n$
message qubits $A_{1,1}\ldots A_{1,n}$, to form the ordered set of
$(n+s)$ qubits $A_{1,1}\ldots A_{1,n+s}$. The $j$-th control qubit
is prepared in $Cstate_1(j)\in \mathcal{S}$ and is in the position
$Cpos_1(j)$, while the relative positions of the $n$ message qubits
are the same as before. Analogously, she forms the set
$A_{2,1}\ldots A_{2,n+s}$.

For $j=1$ to $n+s$, Alice applies the locking operator on qubits
$A_{1,j}A_{2,j}$.

(3) \emph{Communication.} Alice joins the $r$ detect qubits with
$(n+s)$ message and control qubits $A_{1,1}\ldots$ $A_{1,n+s}$, to
form the ordered set of $(n+s+r)$ qubits $A_{1,1}\ldots
A_{1,n+s+r}$. The $j$-th detect qubit is prepared in
$Dstate_1(j)\in \mathcal{S}$ and is in the position $Dpos_1(j)$,
while the relative positions of the remaining $(n+s)$ qubits are the
same as before. Analogously, she forms the set $A_{2,1}\ldots
A_{2,n+s+r}$.

Alice sends $(n+s+r)$ qubits $A_{1,1}\ldots A_{1,n+s+r}$ to Bob, and
$(n+s+r)$ qubits $A_{2,1}\ldots A_{2,n+s+r}$ to Charlie.

Alice waits for Bob and Charlie's acknowledgements of receiving all
the $2(n+s+r)$ qubits. After they have sent their acknowledgements
through unjammable classical communication channel, Alice sends $\{
Dpos_1(j) | \ j = 1,2, \ldots ,r \}$ and the bases of $\{
Dstate_1(j) | \ j = 1,2, \ldots ,r \}$ to Bob. Analogously, Alice
sends $\{ Dpos_2(j) | \ j = 1,2, \ldots ,r \}$ and the bases of $\{
Dstate_2(j) | \ j = 1,2, \ldots ,r \}$ to Charlie.

For each $j = Dpos_1(k)$, Bob
measures qubit $A_{1,j} $ in either $X$ or $Z$ basis, according to
$Dstate_1(k)$, and returns the measurement
result to Alice. If the measurement result is not equal to
$Dstate_1(k)$, Alice announces that a cheating behaviour has been
detected and stops the protocol.

Analogously, for each $j = Dpos_2(k)$, Charlie measures qubit $A_{2,j} $ in either $X$ or $Z$ basis,
according to $Dstate_2(k)$, and returns the
measurement result to Alice. If the measurement result is not equal
to $Dstate_2(k)$, Alice announces that a cheating behaviour has been
detected and stops the protocol.

Now the remaining $(n+s)$ received qubits at Bob's site form the
ordered set $A_{1,1}\ldots A_{1,n+s}$, and the remaining $(n+s)$
received qubits at Charlie's site form the ordered set
$A_{2,1}\ldots A_{2,n+s}$.

 (4) \emph{Unlocking.} For $j=1$ to $n+s$, Bob sends
qubit $A_{1,j}$ to Charlie, and then Charlie returns $A_{1,j}$ to
Bob after performing the unlocking operator on qubits
$A_{1,j}A_{2,j}$ at his site.

If $\exists k, Cpos_1(k)=j$, Bob measures qubit $A_{1,j}$ in either
$X$ or $Z$ basis, according to $Cstate_1(k)$. If the measurement
result does not match $Cstate_1(k)$, he knows that Charlie did not
return the real $A_{1,j}$ and stops the unlocking stage.

Analogously, if $\exists k, Cpos_2(k)=j$, Charlie measures qubit
$A_{2,j}$ in either $X$ or $Z$ basis, according to $Cstate_2(k)$.
If the measurement result does not match $Cstate_2(k)$, he knows
that Bob did not send the real $A_{1,j}$ and stops the unlocking
stage.

Now the remaining $n$ received qubits at Bob's site form the ordered
set $A_{1,1}\ldots A_{1,n}$, and the remaining $n$ received qubits
at Charlie's site form the ordered set $A_{2,1}\ldots A_{2,n}$.

(5) \emph{Decoding.} Bob and Charlie measure qu$N$its $A_1B$ and
qu$N$its $A_2C$ in the Bell basis respectively to achieve
$(b_1,b_2)$ and $(c_1,c_2)$.

\end{document}